# HEALPix View-order for
# 3D Radial Self-Navigated Motion-Corrected ZTE MRI

Curtis A. Corum[1], Stanley J. Kruger[2] and Vincent A. Magnotta[2]
[1]*Champaign Imaging LLC*, [2]*University of Iowa, Department of Radiology*
curt.corum@champaignimaging.com

**Abstract**

*Compressed sensing has reinvigorated the field of non-Cartesian sampling in magnetic resonance imaging (MRI).*

*Until now there has been no 3D radial view-order which meets all the desired characteristics for simultaneous dynamic/high-resolution imaging, such as for self-navigated motion-corrected high resolution neuroimaging.*

*In this work, we examine the use of Hierarchical Equal Area iso-Latitude Pixelization (HEALPix) for generation of three-dimensional (3D) radial view-orders for MRI, and compare to a selection of commonly used 3D view-orders.*

*The resulting trajectories were evaluated through simulation of the point spread function and slanted surface object suitable for modulation transfer function, contrast ratio, and SNR measurement. Results from the HEALPix view-order were compared to Generalized Spiral, 2D Golden Means, and Random view-orders.*

*Finally, we show the first use of the HEALPix view-order to acquire in-vivo brain images.*

## 1. Introduction

Compressed sensing with multi-core and GPU-based reconstruction has renewed and increased interest in 2D, hybrid, and 3D radial non-Cartesian MRI.

View-orders in 3D radial MRI typically cover a spherical (or ellipsoidal) volume of k-space with center-out (spokes) for FID readout as in ZTE, UTE and SWIFT [5] or echo readout through the center (diameters). There are several properties desirable for a 3D radial view-order scheme, not all of which are compatible.

An ideal 3D radial isotropic view-order scheme has all the properties shown in Table 1. Efficiency embodies the favorable properties of equi-distribution, separation, covering, quasi-uniformity and minimal Riesz potential energy described in [3]. Computability is defined as the ability to generate view orders in a deterministic repeatable manner without excessive computation time (such as required for energy minimization). Smooth natural ordering is defined as having a gradient trajectory without large arbitrary jumps or the need for any additional sorting.

Simultaneous self-navigated anatomical/dynamic MRI requires one of the most difficult properties of the view-order scheme. Each subset of views used for a navigator segment must have enough coverage to reconstruct a low resolution image (or provide isotropic spatial information to other estimators) as well as combine into an efficient fully sampled set of view data for anatomical reconstruction.

The Generalized Spiral view-order has been the convenient choice for non-dynamic 3D radial imaging when additional spoiling is not required and is useful for low sound pressure levels in quiet MRI. It posses all the desired characteristics for a 3D radial MRI view-order except the ability to divide into efficient subsets.

The 2D Golden Means view-order [1] has most of the desired characteristics excepting smooth natural order. It is efficient if the number of points in a subset is not too small.

## 2. HEALPix

As seen from Table 1, HEALPix (Hierarchical Equal Area iso-Latitude Pixelization) [2; 3] is a good candidate for a 3D radial dynamic view-order. It possesses all the desirable characteristics of the Generalized Spiral in addition to being divisible into efficient subsets. HEALPix also has the desirable characteristics of the 2D Golden Means view-order, but has smooth natural ordering for the subsets and the whole.

HEALPix was devised by K. M. Górski at the Theoretical Astrophysics Center (TAC) in Copenhagen, Denmark and developed by a dedicated team of collaborators [4]. It was designed with three properties essential for computational efficiency in discretization functions on the unit sphere (S2) and processing large amounts of data:

1. The sphere is hierarchically tessellated into curvilinear quadrilaterals. 2. The pixelization is an equal area partition of S2. 3. The point sets are distributed along fixed lines of latitude.

The last allows for utilization of fast spherical function transforms, and rapid search and addressing. This property may further enable k-space domain estimation of motion.

HEALPix initially divides the sphere into 12 node points with Voronoi regions of equal areas. Each area is then recursively subdivided into smaller equal areas, which for maximal convenience are powers of 2. In the following $N$ is the total number of nodes covering the sphere, which we use as the endpoints of the center-out radial "spoke" views. $N$ is constrained to be $N=12M^2$ where $M$ is a non zero positive integer. For computational efficiency and scaling $M$ is further constrained to be $M=2^K$ where $K$ is a positive integer. The example of $K=2$ is shown in Figure 2. In general, we have the number of nodes: $N=12\times 4^K$

## 3. Methods and Implementation

All generated view-order tables, image data and analysis scripts are deposited at [6] in the interest of reproducible research.

### 3.1. Sequence Modifications

We modified the 3D Radial Silent sequence [7; 8; 8]to accept table input for view-ordering.

### 3.2. View-order Generation

We utilize the specific implementation of HEALPix, "MEALPix" which is Matlab™ based [9]. Additional useful code for HEALPix and other sphere coverage is available from [10]. All

*Table 1: Sphere Coverage Methods*

| Name | Domain | Dynamic Subsets | Computable | Deterministic | Efficiency | Smooth Natural Ordering | References |
|---|---|---|---|---|---|---|---|
| HEALPix | $N = 12k^2$ | ● | ● | ● | ● | ● | |
| 2D Golden Means | $N > 1$ | ● | ● | ● | ● | | Chan/Plews |
| Radial Icosahedral | $N = 10k^2+2$ | ● | ● | ● | ● | | |
| Mesh Icosahedral | $N = 10(j^2+jk+k^2)+2$ | ● | ● | ● | ● | | |
| Cubed Sphere | $N = 6k^2+12k+8$ | ● | ● | ● | ** | | |
| Octahedral | $N = 4k^2+2$ | ● | ● | ● | ** | | |
| Hammersley | $N > 2$ | ● | ● | ● | | | Chan/Plews |
| Random | $N > 1$ | ● | ● | | | | Chan/Plews |
| Generalized Spiral | $N > 2$ | | ● | ● | ● | ● | Wong, Saff |
| Fibonacci | N odd | ● | ● | ● | | | |
| Coulomb* | $N > 1$ | | | | ● | | |

*Log and Max Determinant also, **not equidistant radial views were center-out (spokes). Each consisted of using the unit sphere surface point generated by the MEALPix package and scaled to the gradient discretization value of 32676. The endpoint components were rounded via floor to an integer value.

In addition we generate Spiral [3; 7; 11], 2D Golden and Random – view-order sets [1].

Since HEALPix is constrained to $N = 12 \times 4^K$ views it is not always directly commensurate with the GE Silent view-order scheme of $N = J^2$. For visualization in the figures we use $N = 192$ for HEALPix and $N = 196$ for the Spiral, 2D Golden Means and Random view-orders. For the PSF simulation, Hollow Cube simulation and in-vivo imaging we use $N = 36864$.

Figures 1 and 2 show the gradient waveform shapes and sphere coverage with 192 or 196 views. Each point on the sphere shows the corresponding endpoint of the radial view. Random is not shown.

### 3.3. Gridding Reconstruction

All simulated and acquired datasets were reconstructed using 1.25 oversampled gridding with a radius 2.5 Kaiser-Bessel kernel [14] and 50 iterations of the Pipe-Menon algorithm for sample density correction [15].

### 3.4. Point Spread Function

Simulated point spread functions (PSFs) were generated for each view-order by generating a dataset of k-space values consisting of all ones. Analysis using a custom Octave script was conducted for the properties shown in Table 2. The PSF full width half maximum (FWHM) radius was determined by 3D fitting of a 7-parameter Gaussian model (amplitude, 3 offsets and 3 sigmas).

The maximum value and the volume were determined directly from the data by integrating in a radius 4 neighborhood.

Figure 3 shows slices and a mesh plot of the XY plane in the 32x32x32 neighborhood around the center of the PSF. Scale is logarithmic base 10 to note the rapid falloff.

### 3.5. Modulation Transfer Function

A hollow cube possessing internal slanted sides was simulated by direct synthesis of k-space [12] data and reconstructed by gridding (see Figure 4.) The modulation transfer function (MTF) in an orthogonal XY slice was estimated using the ImageJ plugin Slanted Edge MTF [13]. See Figure 6.

### 3.6. Contrast Ratio and SNR

The contrast ratio and SNR were evaluated using 32x32 voxel patches in the center slice. Black level for contrast came from centering the patch in the dark region at center of the slice. White level consisted of centering the patch on the right of center bright plateau. Noise for the SNR as well as signal were evaluated on the bright patch.

### 3.7. In-vivo Neuroimaging

In-vivo brain images of a healthy human volunteer were taken using the HEALPix view-order and are shown in Figure 5. Images were also taken with the Generalized Spiral view-order.

### 4. Results

As seen in Figure 3 and Table 2 the central region of the PSF for both the Generalized Spiral and HEALPix are similar; HEALPix provides slightly higher residual amplitude around the central point (log10 of -4 vs log10 of -4.5). The fitted FWHM and peak values vary by less than 0.5% percent.

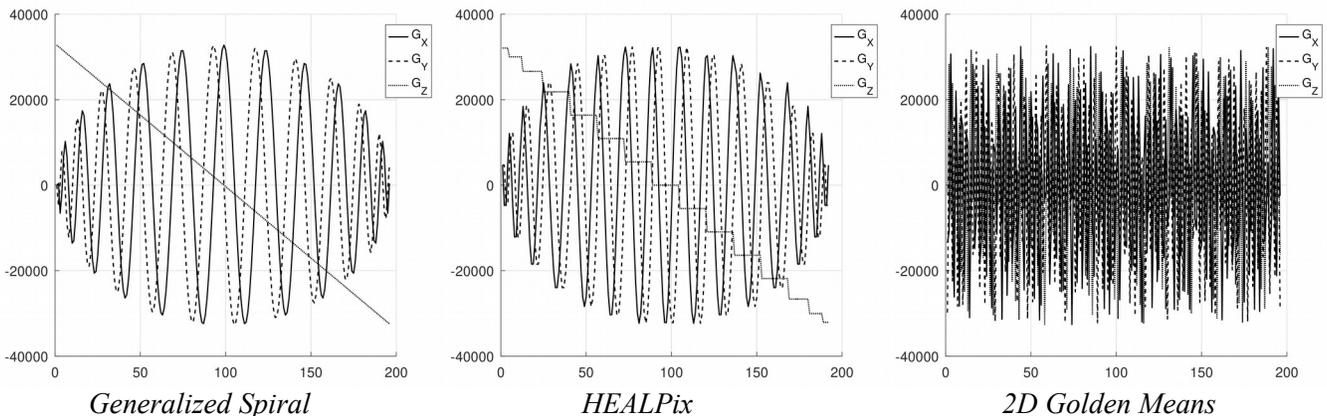

*Figure 1: Gradient Waveforms* – The 2D Golden Means view-order does not have smooth natural ordering.

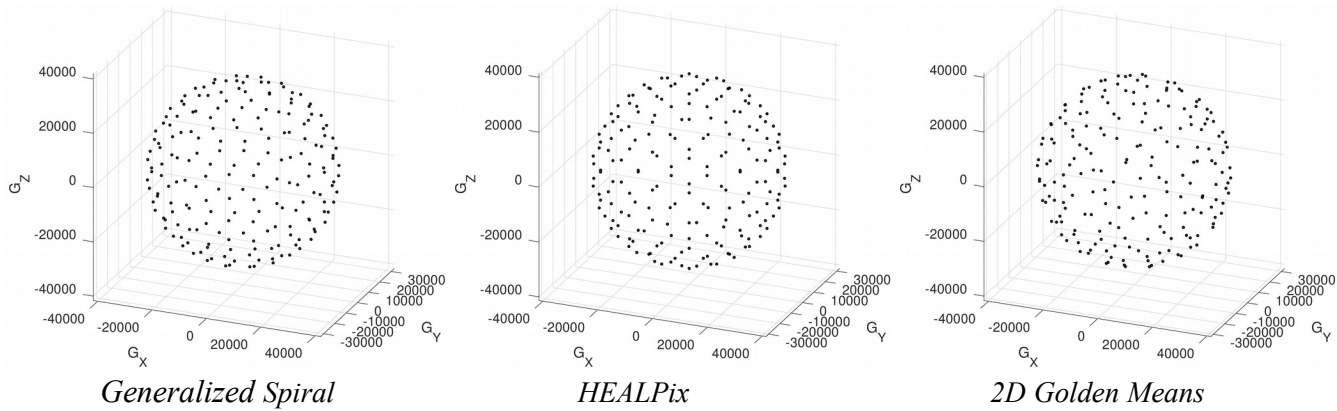

*Figure 2: Sphere Coverage – view endpoints on unit k-space sphere.*

Contrast ratio, MTF performance, and SNR in Table 2 also vary by less than 0.5% between the Generalized Spiral and HEALPix. The values of the 2D Golden Means and Random view-orders fall below the Generalized Spiral and HEALPix.

MTF curve values for HEALPix and Generalized Spiral overlap while 2D Golden Means and Random both fall below for all spatial frequencies in Figure 6.

HEALPix in-vivo brain images (Figure 5) are indistinguishable at automatic optimal window level from those taken with the Generalized Spiral view-order with same number of views (not shown).

## 5. Conclusions

In this initial evaluation, the HEALPix 3D radial view-order possesses all the favorable properties of the Generalized Spiral in addition to being divisible into efficient subsets for dynamic self-navigated protocols. It performs similarly on metrics of PSF, MTF, contrast ratio and SNR as the Generalized Spiral and has yielded initial in-vivo images indistinguishable at automatic optimal window level to images taken using the Generalized Spiral.

## 6. Notes

This work was funded in part by the SBIR Phase I grant R43MH115885 from the NIH/NIMH. Curt Corum and Champaign Imaging LLC are developing technology related to the topics reported in this proceeding. All in-vivo imaging was performed under an IRB-approved protocol at the University of Iowa, Magnetic Resonance Research Facility.

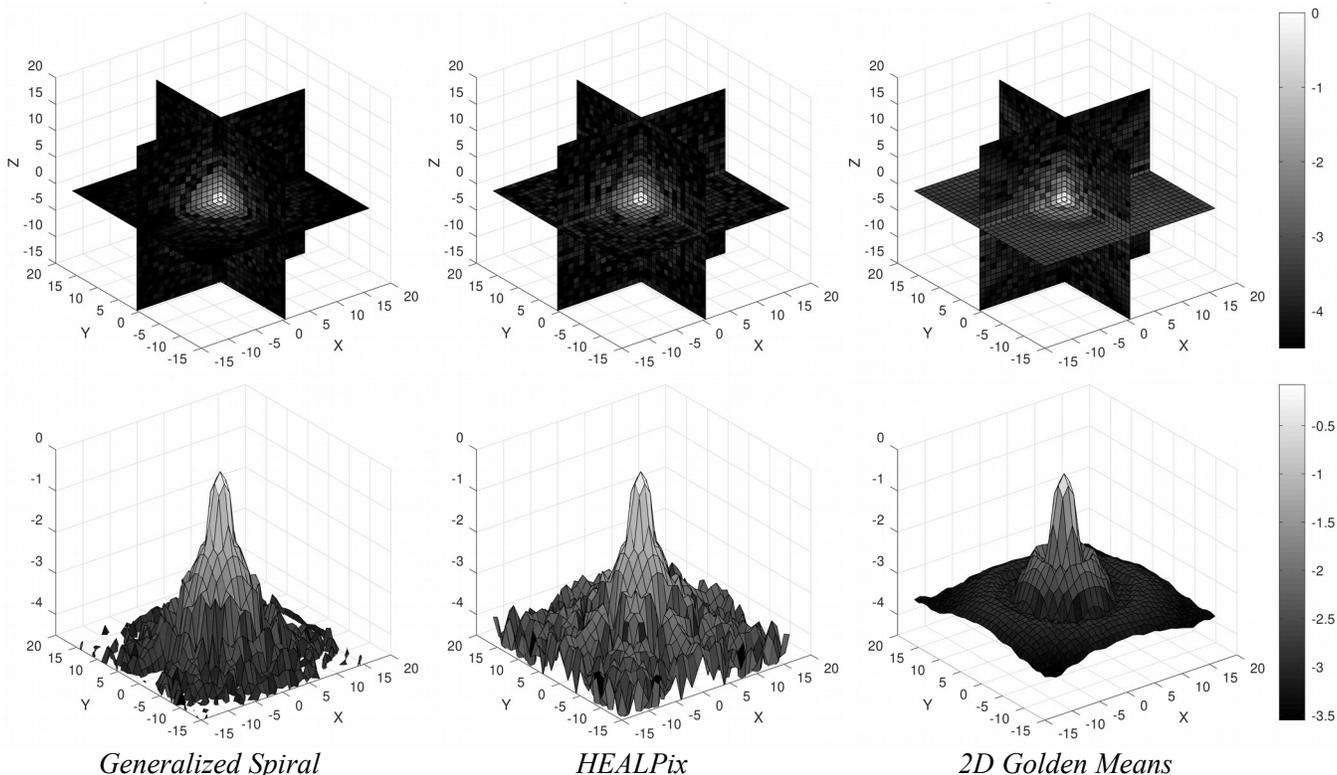

*Figure 3: Log10 of Point Spread Function – orthogonal slices through center (Top), Mesh of XY slices (Bottom).*

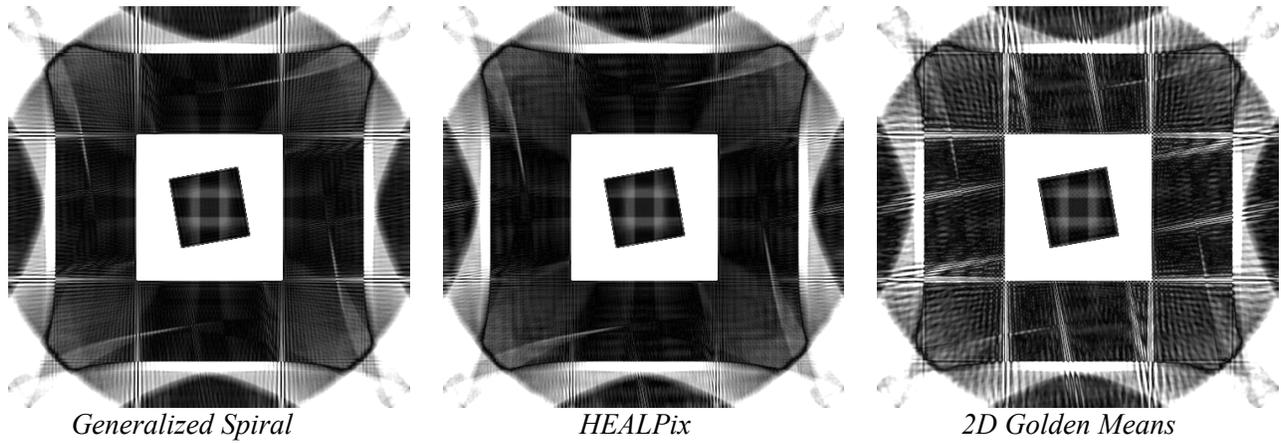

*Figure 4: Simulated Hollow Cube* – for modulation transfer function (MTF), contrast and SNR measurements. Shown at 10x intensity scale and 2.25 nominal field of view (for Nyquist radius).

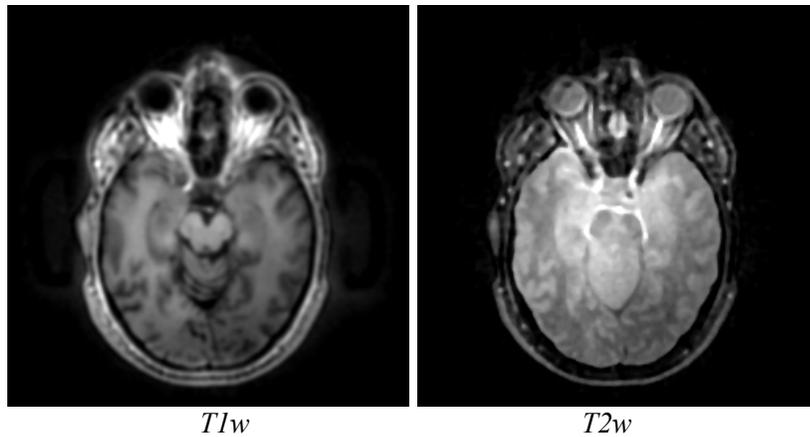

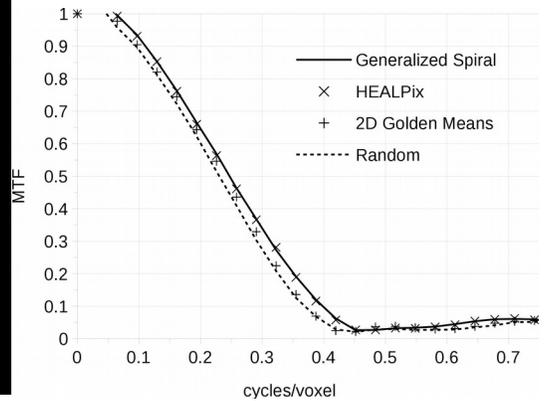

*Figure 5: In-Vivo* - $T_1$ and $T_2$ weighted images with HEALPix view-order.   *Figure 6: MTF Curves* – hollow cube slanted edge,

**Table 2: View-order Performance**

| Name | Views | 3D matrix | PSF Peak | Volume [voxels] | FWHM [voxels] | Contrast | 50% MTF [cycles/voxel] | SNR |
|---|---|---|---|---|---|---|---|---|
| Generalized Spiral | 36864 | $192^3$ | 0.93625 | 13.034 | 2.070 | 66.73:1 | 0.2470 | 166.83 |
| HEALPix | 36864 | $192^3$ | 0.93617 | 13.142 | 2.073 | 67.73:1 | 0.2459 | 166.83 |
| 2D Golden Means | 36864 | $192^3$ | 0.83892 | 12.114 | 2.075 | 58.88:1 | 0.2393 | 142.86 |
| Random | 36864 | $192^3$ | 0.70754 | 13.038 | 2.146 | 66.80:1 | 0.2302 | 125.13 |